\documentclass[lettersize,journal,10pt]{IEEEtran}
\usepackage{amsmath,amsfonts}
\usepackage{hyperref}
\hypersetup{
	colorlinks=true,
	linkcolor=black,
	citecolor=blue,
	urlcolor=blue
}
\usepackage{cite}

\usepackage{amsthm}  % 定义定理环境
\usepackage{amsmath} % 数学公式
\usepackage{amssymb} % 数学符号（如\square）
\usepackage{svg}

\usepackage{booktabs}
\usepackage{colortbl}
\definecolor{lightgray}{gray}{0.95}
\usepackage{svg}

\usepackage[linesnumbered,boxed,ruled]{algorithm2e}

\usepackage{array}
\usepackage{amsmath}
\usepackage{textcomp}
\usepackage{stfloats}
\usepackage{url}
\usepackage{verbatim}
\usepackage{graphicx}
\usepackage{tabularx}
\usepackage{cite}
\usepackage{cleveref}
\usepackage{color}
\usepackage{svg}
\usepackage{soul} % 导入 soul 包，用于高亮显示
\usepackage{graphicx}
\usepackage{subcaption} % 必须包含此宏包以支持 subfigure 环境
\usepackage{caption}

\usepackage{ulem}

\usepackage{graphicx}
\usepackage{subcaption}

\def\BibTeX{{\rm B\kern-.05em{\sc i\kern-.025em b}\kern-.08em
    T\kern-.1667em\lower.7ex\hbox{E}\kern-.125emX}}
\begin{document}

\title{Token Communications (TokCom): A Unified AI-Native Communication Framework}
%Token Communications: Architecting the Agentic Era of 6G and AI-Native Networks}

    \author{
       { Yaru Fu,~\IEEEmembership{Member,~IEEE}, Liang Ji,~\IEEEmembership{Student Member,~IEEE}\\Sabita Maharjan,~\IEEEmembership{Senior Member,~IEEE},~and Tony Q. S. Quek,~\IEEEmembership{Fellow,~IEEE} }
      % \\ Yue Gao,~\IEEEmembership{Fellow,~IEEE},~and Tony Q. S. Quek,~\IEEEmembership{Fellow,~IEEE} }
       %Ricky Y. K. Kwok,~\IEEEmembership{Fellow,~IEEE},~ 

\thanks{This work was supported in part by the Team-based Research Fund under Reference No. TBRF/2024/1.10, in part by the Research Matching Grant under Reference No. CP/2025/1.1, in part by the grant from the Research Grant Council (RGC) of the Hong Kong Special Administrative Region, China,
under Project Reference No. UGC/FDS16/E06/25, and in part by the National Research Foundation, Singapore and Infocomm Media Development Authority under its Communications and Connectivity Bridging Funding Initiative. Any opinions, findings and conclusions or recommendations expressed in this material are those of the author(s) and do not reflect the views of the National Research Foundation, Singapore.
%\textit{(Corresponding author: Yaru Fu)}

Y. Fu and L. Ji are with the School of Science and Technology,  Hong Kong Metropolitan University, Hong Kong, 999077, China (e-mail: yfu@hkmu.edu.hk, lji@hkmu.edu.hk).

  %Yan Kyaw Tun is with the Department of Electronic Systems, Aalborg   University, 2450 København, Denmark (e-mail: ykt@es.aau.dk).

  S. Maharjan is with the Department of Informatics, University of Oslo, Oslo, Norway (e-mail: sabita@ifi.uio.no).

  %Y. Gao is with the Institute of Space Internet, Fudan University, Shanghai 200438, China (e-mail: gao.yue@fudan.edu.cn).

  Tony Q. S. Quek is with the Department of Information Systems Technology and Design, Singapore University of Technology and Design, Singapore 487372 (e-mail: tonyquek@sutd.edu.sg). 
}
}
\maketitle

\begin{abstract}
As artificial intelligence (AI) evolves from static perception to generative reasoning and autonomous agency, the fundamental principles of wireless communications are undergoing a paradigm shift. The classical Shannon paradigm, centered on reliable bit-level reconstruction for users, is increasingly misaligned with an emerging scenario in which the primary users of the network are interconnected AI agents. This article introduces token communications (TokCom), a novel framework that elevates tokens, i.e., the fundamental processing units of large language models (LLMs), to first-class entities for information exchange in the sixth generation wireless cellular networks (6G). We first examine the architectural transition from conventional communication systems to TokCom and identify the key challenges in implementing this transition, along with potential solution approaches. Thereafter, we present a practical case study to demonstrate the effectiveness of token sharing among heterogeneous language models. Finally, we outline promising future research directions toward realizing an AI-native, token-driven communication paradigm suitable for 6G.
\end{abstract}

\section{Introduction}
The evolution of wireless cellular communications has been driven primarily by the pursuit of higher data rates to satisfy ever-increasing human sensory demands \cite{6G-wifi7-wifi8}. From the narrowband voice services of first-generation wireless networks (1G) to the immersive 4K video capabilities of fifth-generation communication systems (5G), the fundamental optimization objective has remained largely unchanged, i.e., achieving accurate, high-fidelity reconstruction of source data at the receiver. However, the emerging vision for 6G points to a substantially different paradigm, in which the network is expected to natively support a wide range of intelligent applications \cite{AI-6G}. In other words, we are entering an era characterized by agentic artificial intelligence (AI), in which autonomous entities are capable of reasoning, planning, and executing complex tasks collaboratively. In this context, the user of the network is no longer solely a human operating a handheld device, but increasingly an AI agent interacting with other agents, edge nodes, and cloud infrastructures, thereby imposing fundamentally new requirements on the design of future wireless cellular networks.

The foregoing shift calls for a departure from traditional bit-level-accurate transmission. In conventional wireless systems, the objective is to reproduce the source signal at the receiver with minimum distortion, without considering how that information is ultimately used \cite{5G-6G}. However, for an AI agent tasked with decision-making or control, raw video frames or speech waveforms are often highly redundant intermediate representations. The agent does not act directly on individual pixels or waveform samples; instead, it operates on abstracted concepts such as object identities, scene layouts, action intents, or dialog states \cite{GAI-Com}. From this perspective, enforcing bit-accurate reconstruction of sensor data can be wasteful in both spectrum and energy, as it optimizes for human perception rather than machine utility.
What such agents primarily require is the exchange of high-level knowledge, context, and intent that is directly relevant to their tasks. This observation is consistent with the motivation behind semantic communications (SemCom), which has laid an important foundation by emphasizing the transmission of meaning rather than raw data. Nonetheless, most existing SemCom approaches \cite{GAI-Com,Multi-Modal-SemCom,Token-MLLM} are tightly coupled to specific tasks or modalities, relying on pre-designed encoders that extract task-dependent features. This design often limits generalizability, complicates interoperability, and makes it difficult to share or reuse learned representations across heterogeneous applications and intelligent devices \cite{Token_Review,Protocols_Security,AI_Security}.

Token communications (TokCom) is proposed as a practical realization of semantic communication principles tailored to the era of large language models (LLMs). In LLMs and multimodal LLMs, tokens serve as a common abstraction layer that maps diverse inputs such as text, images, audio, and even structured sensor data into a unified, high-dimensional embedding space. By treating these tokens as a universal representation of machine intelligence, TokCom introduces a standardized, modality-agnostic unit of exchange that is already well aligned with state-of-the-art AI models. This yields several concrete benefits: (i) it allows communication protocols to directly interface with pretrained foundation models without relying on feature engineering; (ii) it supports cross-modal fusion and reasoning, since different data types are expressed in a shared token space; and (iii) it enables task-oriented compression, where only the most informative tokens for inference or control need to be transmitted.

Under this paradigm, the network is no longer a passive bit pipe but becomes tightly coupled with computation, effectively operating as a distributed neural processing substrate. Tokens generated by one agent can be interpreted, transformed, or extended by other agents and edge/cloud models using the same underlying token semantics. This integration reduces end-to-end latency (by avoiding redundant decoding and re-encoding at each hop), improves robustness (through generative reconstruction of missing tokens), and enhances resource efficiency (by prioritizing tokens that maximally impact downstream decisions). As a result, TokCom provides a technically grounded and scalable pathway for implementing AI-native, semantics-aware communication in 6G networks.

However, the realization of TokCom also faces several challenges, including the design of token-aware network architectures, the integration of LLM-based processing into existing protocol stacks, and the guarantee of reliability, scalability, and security in distributed AI settings. In this article, we first present a potential end-to-end architecture for TokCom, highlighting how tokens can be incorporated as first-class entities in 6G systems. We then discuss the key technical issues and possible solutions, such as token-level resource management, generative reliability, and context-aware networking. Next, we present a case study to illustrate the effectiveness of TokCom in a representative multi-agent collaboration scenario. Finally, we conclude the article by summarizing the main insights and outlining promising directions for future research on AI-native, token-driven communications.

\begin{figure*}[t] 
    \centering
    \includegraphics[width=0.860\textwidth]{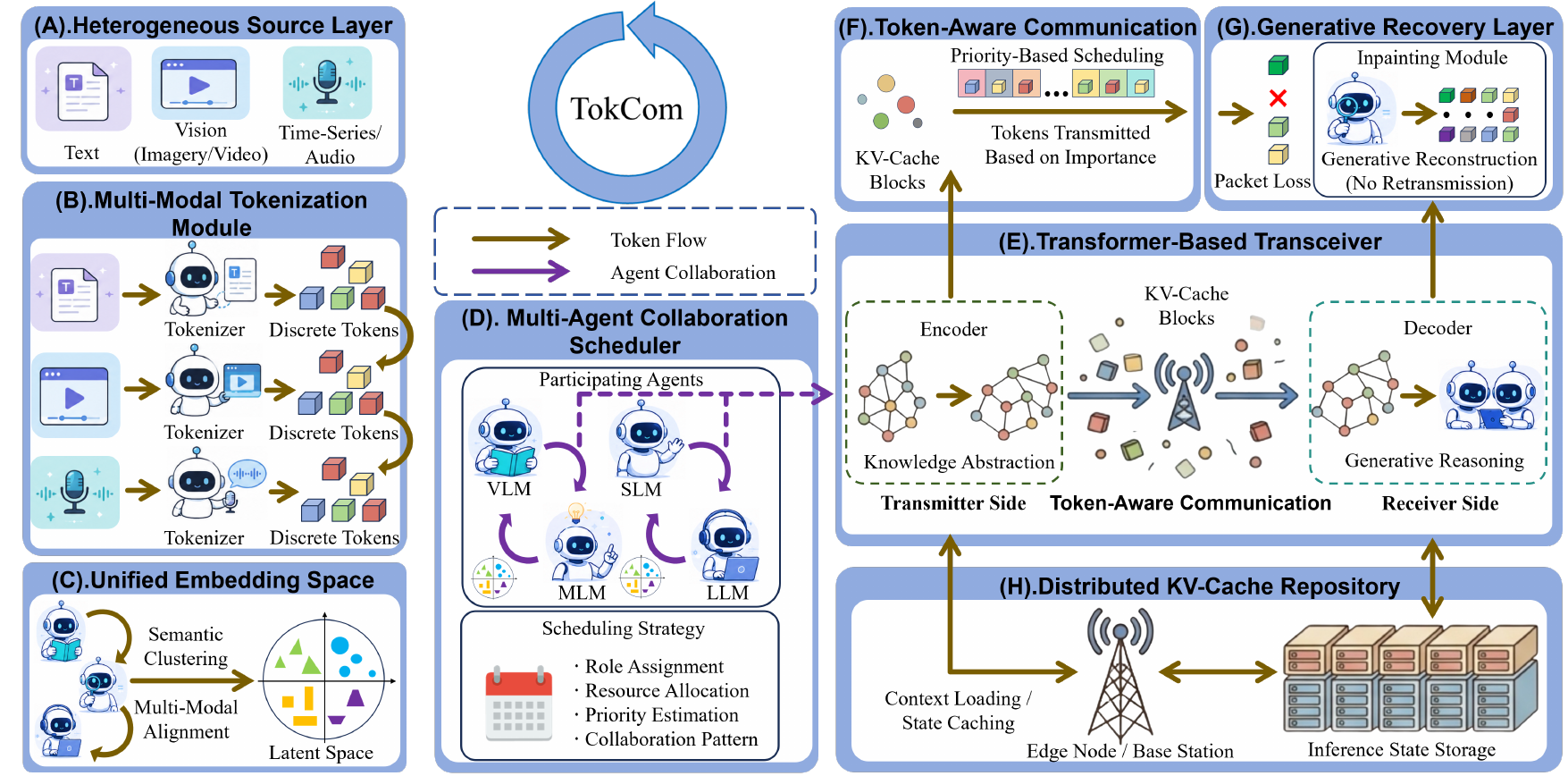}
    \caption{A high-level illustration of the TokCom framework.}
    \label{fig:TokCom}
\end{figure*}

% (A) Multi-modal data streams of text, vision, and audio originating from the physical world. (B) Mapping modality-specific signals into discrete semantic tokens, replacing conventional source coding. (C) Aligning tokens from different modalities in a shared latent space for cross-modal semantic consistency. (D) Knowledge abstraction at the encoder and generative reasoning at the decoder, replacing the physical-layer protocol stack. (E) Transmitting semantically meaningful tokens with priority-based scheduling rather than raw bits. (F) Reconstructing lost tokens via generative inpainting to ensure reliability without retransmission. (G) Edge caching of inference states to reduce redundant transmissions and enable low-latency context switching.

\section {Architecture, Key Issues, and Potential Solutions}
In this section, we first elaborate on the architectural structure of TokCom. We then discuss the key technical challenges associated with its practical implementation, followed by an overview of potential solution approaches and design guidelines.
\subsection{Architecture}

To fully realize the potential of TokCom, the 6G architecture must be redesigned to be token-aware rather than merely bit-aware. As illustrated in Fig. \ref{fig:TokCom}, in a TokCom-enabled system, the end-to-end pipeline begins with the heterogeneous source layer, where multi-modal data streams such as text, vision, audio, and sensor signals are generated from the physical world. These heterogeneous inputs are first processed by multi-modal tokenization module, which maps modality-specific signals into discrete semantic tokens, thereby replacing conventional source coding. The resulting tokens are then projected into a unified embedding space, where representations from different modalities are aligned to preserve cross-modal semantic consistency.

TokCom incorporates a multi-agent collaboration scheduler to support collaborative intelligence. This module selects the participating agents (e.g., LLMs and Small Language Models (SLMs)) and performs role assignment as well as other scheduling strategies. To capture contextual dependencies across token streams, the transceiver adopts a transformer architecture, and the token streams are processed by a communication encoder that prepares them for transmission over the wireless channel, taking into account both radio link conditions and task-level priorities. The token-aware communication module transmits semantically meaningful tokens rather than raw bits, so that wireless resources can be allocated according to task importance and semantic value. This tight coupling between token generation and radio interface design is essential; namely, the network is no longer transporting raw samples for human interpretation, but structured units that directly drive downstream AI reasoning.

The most transformative element of this architecture is the generative recovery layer. In traditional communication systems, packet loss is handled primarily through rigid retransmission protocols such as automatic repeat request, which aim to restore bit-level correctness \cite{Repeat_Request_Com}. In contrast, TokCom introduces the notion of generative reliability, where the goal is to preserve task performance rather than exact symbol recovery. When a portion of the token stream is lost due to channel fading or interference, the receiver does not necessarily trigger an immediate retransmission. Instead, it invokes a local generative model, i.e., the inpainter, to infer the missing tokens from the surrounding token context and its internal priors. This process is analogous to how humans can reliably understand a sentence even when some words are muffled or missing, by leveraging linguistic structure and semantic expectations.

By exploiting the predictive capabilities of modern foundation models, TokCom can maintain high task-level or collaboration performance (e.g., in reasoning, control, or coordination) even in lossy or interference-limited environments. In many AI-native applications, slight deviations in intermediate tokens that do not affect the final decision are acceptable, making strict bit-perfect delivery unnecessarily costly \cite{Token_Exit}. Generative reliability therefore allows the system to trade a small amount of symbol-level accuracy for significant gains in spectral efficiency, latency, and robustness. Conceptually, this shifts part of the reliability burden from the physical and link layers to an intelligence layer, where errors are repaired or compensated using learned world models, enabling 6G systems to operate effectively in regimes that would traditionally be considered too noisy for reliable communication.

Another critical enabler is distributed Key-Value (KV) cache repository. In LLM inference, the KV cache stores the intermediate attention states associated with previously processed tokens, thereby accelerating the generation of subsequent tokens and preserving conversational or reasoning context \cite{KV-Cache_Spillover}. In a distributed TokCom system, the network itself can act as a distributed KV cache. Edge nodes in an AI-enabled radio access network (AI-RAN) can maintain the KV caches of ongoing agentic sessions, allowing mobile or distributed AI agents to hand off their reasoning state as they move across cells or offload computation between edge and cloud. 
This form of context-aware networking ensures that AI agents do not need to restart their reasoning or dialog from scratch after each handover or offload event, significantly reducing both latency and computational redundancy. More broadly, it marks a conceptual shift from networks that simply forward packets to networks that can remember, predict, and actively assist in the execution of intelligent tasks. Such a token-aware, cache-enabled architecture provides a concrete and technically grounded pathway for integrating communication and computation in future 6G systems.

\subsection{Key Issues and Potential Solutions}
\subsubsection{Semantic Integrity and the Hallucination Propagation Problem} In traditional digital communication, failure is usually measured by the bit error rate. If a bit is erroneous, the checksum does not match, and the packet is either corrected or discarded. In TokCom, the more serious problem is not an error bit, but a semantic hallucination \cite{Sematic_Hallucination}. The Token Inpainter at the receiver may fill in a missing token that is grammatically correct but factually or logically wrong, for example, turning a negative command into a positive one in a robotic control sequence. In this case, the connection still works, but the meaning is no longer trustworthy. The danger grows in multi-agent workflows, where the output of one AI system becomes the input of the next, allowing small errors to spread and snowball across the network.

To reduce this risk, token-level verification protocols become essential. Instead of just checking bits, TokCom needs a semantic checksum. One idea is to send small anchor embeddings, compact numerical summaries of the intended meaning, together with the token stream. The receiver can then compare its predicted or recovered tokens with these anchors to ensure that the overall meaning remains on a reliable path. In addition, cross-model consensus can be used; for instance, several small models at the edge independently judge critical tokens and take a majority vote. As demonstrated in \cite{Vote_LLM}, this lightweight voting scheme provides a practical safeguard against drift, hallucination, and unreliable behaviors in generative and multi-LLM systems.

\subsubsection{KV-Cache Explosion} Tokens are an efficient abstraction for efficiently representing and communicating information, but the mechanism that enables token-based processing, namely, the KV cache, introduces a substantial resource burden. To maintain conversational or task context, an AI agent must store or exchange the attention states associated with all preceding tokens. As the context window of state-of-the-art models scales from thousands to millions of tokens, the KV cache can grow to several gigabytes, which far exceeds the practical memory and bandwidth budgets of typical 6G systems \cite{KV-Cache_Spillover}. This leads to a new type of memory wall, as transferring context over the network can take longer than performing inference itself.

To address this bottleneck, several directions can be explored. One is dynamic KV-cache orchestration with pipeline overlapping \cite{KV-Cache_Spillover}. In such schemes, KV cache is dynamically managed across heterogeneous memory hierarchies (e.g., GPU and CPU), where attention states with limited contribution to the current inference step are selectively offloaded or recomputed to balance memory pressure and execution efficiency. From an edge-network perspective, AI-native edge context management will be a promising enabler.
In this paradigm, 6G edge nodes (e.g., access points and base stations) act not only as communication relays but also as context repositories. When an AI agent moves from one cell to another, the network performs a context handover, pre-loading the agent’s KV cache onto the target edge node. This avoids transmitting the full history over the air and effectively turns the wireless infrastructure into a distributed, persistent memory substrate that follows mobile AI agents. This architecture naturally fits within a cloud-edge-device collaborative framework, enabling flexible partitioning of context and computation to balance latency, bandwidth, energy, and compute resources across the system.
\subsubsection{Non-deterministic Latency and Token-Native Scheduling} Existing wireless cellular networks are primarily designed for quasi-continuous (fluid) data streams, where packet arrivals are relatively stable and predictable. TokCom traffic, in contrast, is inherently bursty and non-deterministic. For example, a LLM may emit several tokens almost instantaneously and then pause for on the order of 200 ms to perform internal computation before generating the next token \cite{Token-MLLM}. Conventional scheduling algorithms (e.g., proportional fair scheduling) are not well matched to such behavior, and may result in buffer build-up, underutilized resource blocks, or both. In addition, for TokCom, the key latency metric shifts from time-to-first-byte to time-to-first-decision, i.e., the time until a useful control or semantic action can be taken.

Addressing this mismatch requires token-aware radio resource management. Future TokCom-enabled 6G systems must be able to account for the semantic priority of token streams when allocating resources. For example, in an autonomous driving scenario, tokens associated with obstacle detection should bypass standard queues and be assigned pre-determined resource blocks, whereas tokens conveying less time-critical information, such as environmental description, can be deferred. We further anticipate the development of predictive scheduling mechanisms, in which the edge nodes execute lightweight language models to estimate the length and timing of upcoming token bursts based on the current interaction state. This prediction will depend on sufficiently high forecasting accuracy, as the benefit of resource pre-allocation must be balanced against the computation overhead on edge devices. By forecasting the AI agent’s traffic patterns, the network can pre-allocate resources more efficiently, thereby smoothing the inherently unpredictable traffic of generative models into a more stable, predictable flow.

\subsubsection{Joint Resource Management for TokCom}
Although the non-deterministic scheduling in \textit{3)} primarily addresses the temporal domain of transmission, it directly translates into spatially multidimensional resource-allocation scaling challenges across the network.
%The transition to TokCom fundamentally changes how spectral, computational, and storage resources are managed in 6G networks. 
Conventional resource allocation aims to balance throughput and latency for bit-accurate data streams. %In contrast, resource demand in a TokCom-enabled network is tightly coupled to the instantaneous intelligence requirements of AI agents. 
In contrast, multi-tiered networks operating under TokCom demand a multi-resource coupled approach where spectral, computational (inference FLOPS), and storage capabilities must be jointly optimized based on the tokenized task requirements.
A key challenge is the coupled resource heterogeneity, where an agent may generate low-rate status tokens for long periods, then abruptly require substantial computational and spectral resources for complex reasoning or multimodal draft-and-verify cycles. Static or purely traffic-driven schemes are unable to account for this computational-communicational coupling, leading to reasoning bottlenecks or inefficient resource utilization.

To tackle this problem, intelligence-aware resource orchestration (IARO) is essential to dynamically manage the multi-type resources. In IARO, the network leverages a multi-tier scheduling framework that jointly manages inference FLOPS, token bandwidth, and context cache as coupled resources. The edge nodes can employ reinforcement learning-based slicing to dynamically create token slices with differentiated priorities based on the semantic criticality of the underlying task. For example, tokens associated with safety-critical embodied AI functions (such as collision avoidance) are mapped to a high-priority reasoning slice with guaranteed low-latency compute-and-forward capabilities. In addition, predictive resource pre-emption may be realized by lightweight SLMs that estimate ongoing token trajectories. When the predictions are sufficiently accurate, the system can pre-activate edge computing or reserve bandwidth ahead of a forthcoming reasoning burst. This shift from reactive, traffic-centric control to proactive, intelligence-aware orchestration enables 6G networks to elastically match their resources to the time-varying resource requirements of distributed AI agents.

\begin{figure*}[t]
    \centering
    \begin{subfigure}[b]{0.8\textwidth}
        \centering
        \includegraphics[width=\linewidth,height=\linewidth,keepaspectratio]{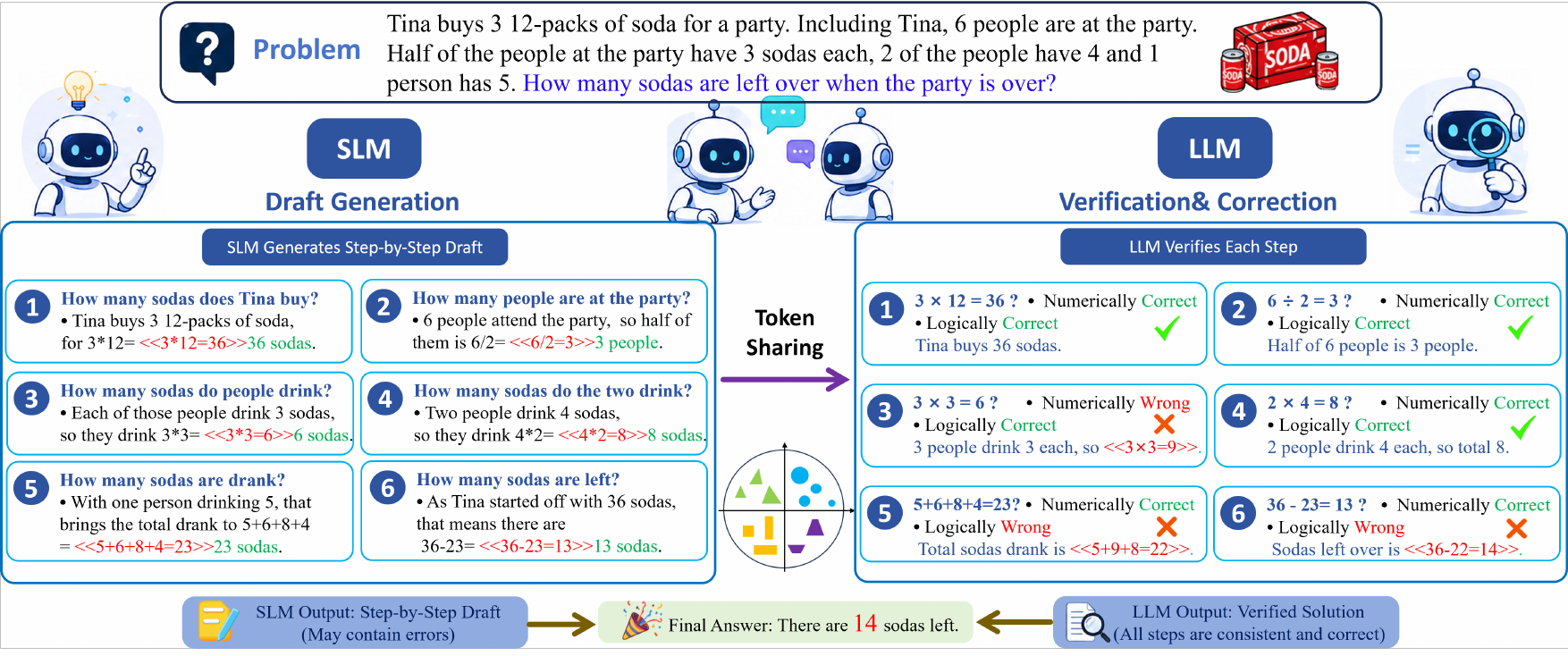}
        \caption{}
        \label{Vision}
    \end{subfigure}

    \begin{minipage}{\textwidth}
        \centering
        \begin{subfigure}[b]{0.39\textwidth}
            \centering
            \includegraphics[width=\linewidth,height=\linewidth,keepaspectratio]{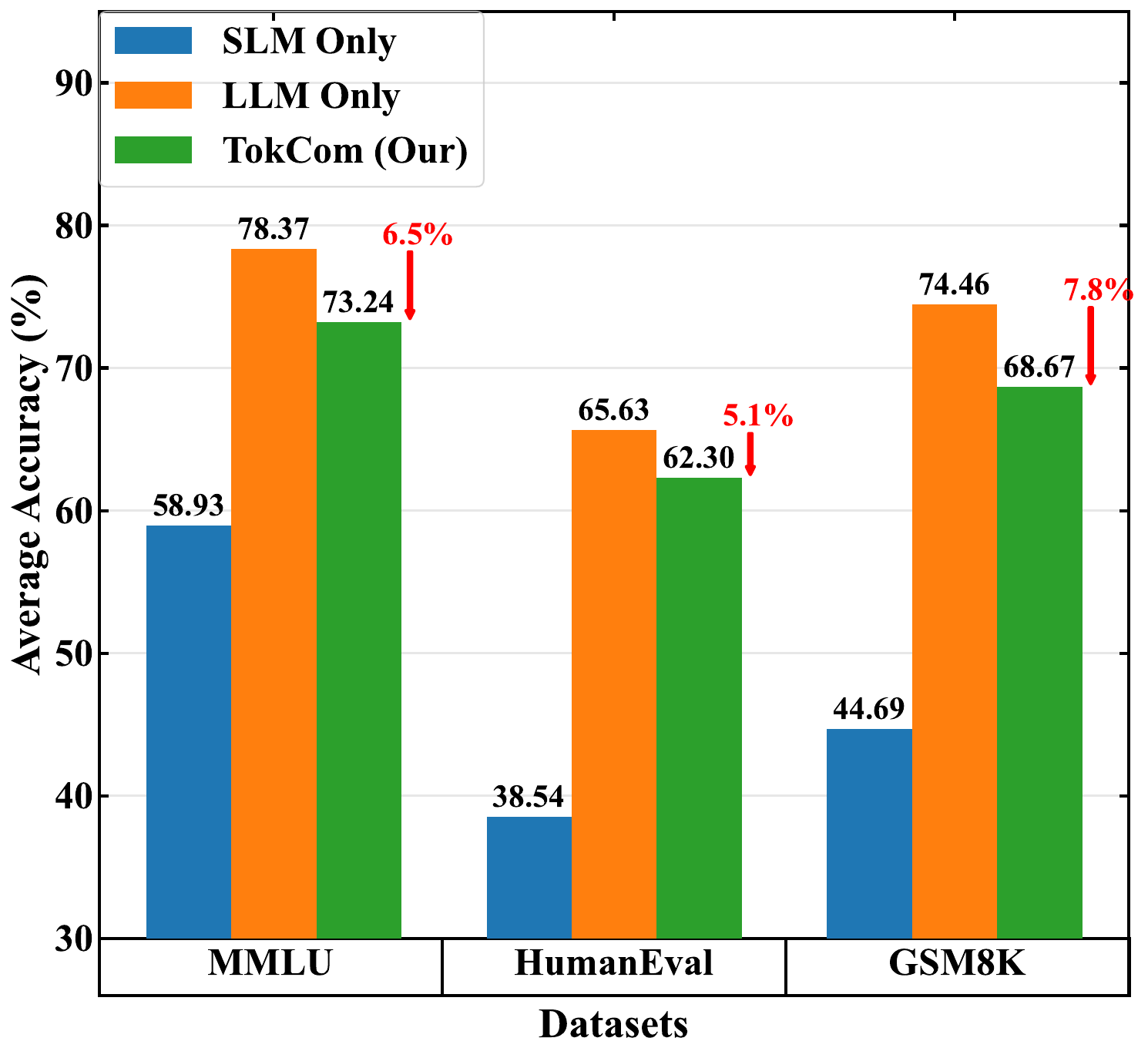}
            \caption{}
            \label{OA}
        \end{subfigure}
        \hspace{0.02\textwidth}
        \begin{subfigure}[b]{0.45\textwidth}
            \centering
            \includegraphics[width=\linewidth,height=\linewidth,keepaspectratio]{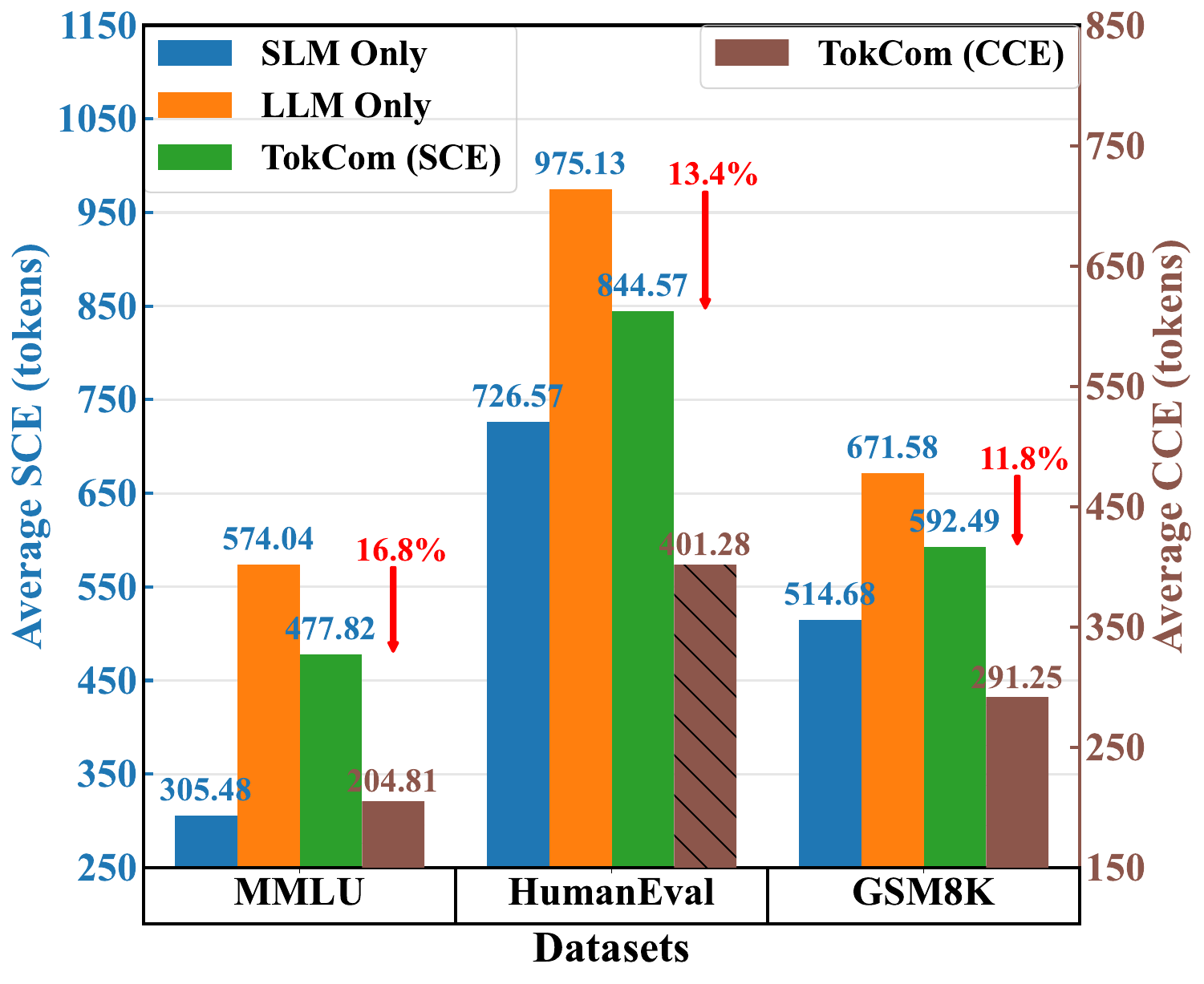}
            \caption{}
            \label{CCE+SCE}
        \end{subfigure}
    \end{minipage}

    \caption{Performance evaluation of TokCom: (a) Visualization of multi-agent collaboration; (b) Overall accuracy; (c) System Computational Efficiency (SCE) and Collaborative Communication Efficiency (CCE), both of which are lower is better.}
    \label{Experiment}
\end{figure*}

\section{Case Study}
The TokCom framework can play a crucial role in enabling collaboration among AI agents. In this section, we present a case study on token-level communication between LLM agents, focusing on inference quality and system efficiency. This case study illustrates how TokCom enhances cooperative inference, offering a practical solution for efficient multi-agent collaboration.

\subsection{Experimental Setup}

In our case study, we employ an SLM-based agent with fewer than 10 billion parameters and an LLM-based agent with more than 10 billion parameters, each assigned a distinct role  \cite{LLM_Survey}. Specifically, the SLM agent is instantiated with the lightweight Vicuna v1.5-7B model, whereas the LLM agent is realized using the more powerful Llama 2-13B model. Both models share an identical tokenizer with a vocabulary of 32,000 tokens, thereby enabling direct token-level communication without any intermediate alignment. Multi-agent collaboration is realized at the token-sharing level. Concretely, the input prompt is first tokenized by the shared tokenizer and fed into the SLM, which performs inference and produces draft reasoning tokens that capture the logical skeleton but may contain errors. This token sequence is then directly transmitted to the LLM. The LLM concatenates these tokens with the original prompt tokens, verifies and corrects potential errors, and finally generates the refined answer. In Fig. \ref{Experiment}(a), we illustrate the collaborative process between the SLM and LLM agents. The SLM agent is responsible for generating a step-by-step draft solution, while the LLM agent verifies and refines each step within the same token space.  Owing to its relatively small parameter size (i.e., 7B), the SLM agent offers weaker reasoning capability but lower inference overhead. In contrast, the LLM agent, with a larger parameter size (i.e., 13B), provides stronger reasoning ability at the cost of higher inference latency, making it well-suited for error correction and the generation of high-quality final answers.

In our experiment, we consider three multi-agent collaborative reasoning tasks, namely broad knowledge understanding, code generation, and mathematical reasoning. These tasks are evaluated on three benchmark datasets: Massive Multitask Language Understanding (MMLU), Human Evaluation (HumanEval), and Grade School Math 8K (GSM8K), respectively. In particular, MMLU is a large-scale multi-task benchmark consisting of approximately 15,900 multiple-choice questions spanning 57 subjects across the humanities, social sciences, and STEM disciplines; we randomly select 2,000 problems, i.e., $N=2000$. HumanEval is a benchmark dataset containing 164 hand-crafted Python programming problems (i.e., $N=164$) with unit tests for evaluating functional code generation. GSM8K is a benchmark dataset consisting of around 8,500 grade-school math word problems, from which we randomly select 2,000 problems, i.e., $N=2000$; each problem requires multi-step reasoning to derive a final numerical answer.
To comprehensively evaluate the effectiveness of the proposed TokCom scheme, we compare it with the following two baseline methods: \textbf{1) SLM-Only}: Vicuna v1.5-7B independently solves each problem and directly generates the final answer without any assistance from the LLM agent; and  \textbf{2) LLM-Only}: Llama 2-13B performs end-to-end reasoning independently based on its self-generated step-by-step reasoning draft.

We use the following three primary metrics to assess the reasoning performance, communication efficiency, and computational overhead:
\begin{itemize}
    \item Overall Accuracy (OA): The proportion of correctly answered problems, defined as $\mathrm{OA} = N_\mathrm{cor} / {N}$, where $N_\mathrm{cor}$ denotes the number of correctly answered problems.
    \item Collaborative Communication Efficiency (CCE): The average number of tokens sent from SLM to LLM per correctly answered problem, defined as $\mathrm{CCE} = \left( \sum_{i=1}^{N_\mathrm{cor}} C_{S,i} \right) / {N_\mathrm{cor}}$, where $C_{S,i}$ is the number of tokens generated by the SLM at the decoding stage for the $i$-th correctly answered problem.
    \item System Computational Efficiency (SCE): The average total number of tokens generated by both agents per correctly answered problem, defined as $\mathrm{SCE} = \sum_{i=1}^{N_\mathrm{cor}} \left( C_{S,i} + C_{L,i}\right) / {N_\mathrm{cor}}$, where $C_{L,i}$ denotes the number of tokens generated by LLM for the $i$-th correctly answered problem.
\end{itemize}
Lower CCE and SCE values indicate higher communication and computational efficiency, respectively. Together, these metrics capture the token-level communication overhead and the system-level computational cost. All reported results are averaged over five independent runs to ensure reliability and robustness.

\subsection{Performance Analysis}
To better illustrate the effectiveness of collaboration (TokCom) among different agents, we use the 60th problem in GSM8K as a representative example, as shown in Fig. \ref{Experiment}(a). The draft generated by the SLM may contain some errors, including numerical miscalculations and logical mistakes. For example, in Step 3, the SLM incorrectly computes $\ll3\times3=6\gg$. In Step 5, it mistakenly sums the number of sodas with the number of people. These mistakes lead to an incorrect final result. After receiving the SLM draft, the LLM verifies and revises it step by step, correcting these errors and ensuring an accurate final answer. This example clearly shows that TokCom can leverage the low-cost drafting capability of the SLM and the stronger verification ability of the LLM to achieve more reliable collaborative reasoning. Subsequently, we analyze the performance of our proposed scheme against the baselines in terms of the previously defined evaluation metrics.

In Fig. \ref{Experiment}(b), we compare the overall accuracy across the three datasets. Among all methods, SLM-Only achieves the lowest performance, with an accuracy of only 38.54\% in HumanEval, indicating that the SLM has limited reasoning capability, especially for tasks requiring precise program synthesis and multi-step inference. In contrast, LLM-Only achieves the highest accuracy, reaching 65.63\%, which demonstrates its stronger reasoning ability and better capacity for handling complex tasks. Our proposed TokCom achieves accuracy close to that of LLM-Only, with gaps of only 6.5\%, 5.1\%, and 7.8\% on the three datasets, respectively. The reason is that in TokCom, the LLM verifies and revises the draft produced by the SLM, thereby combining the SLM’s drafting capability with the LLM’s stronger verification ability to obtain reliable results. This approach is efficient, but it comes with a slight reduction in accuracy.

In Fig. \ref{Experiment}(c), we evaluate the system computational efficiency of different schemes. It can be seen that LLM-Only incurs the highest computational cost, far exceeding that of SLM-Only. For example, on MMLU, each correctly answered problem requires an average of 574 tokens for LLM-Only, whereas SLM-Only requires only 305 tokens. In addition, the proposed TokCom significantly reduces the cost compared with LLM-Only by 16.8\%, 13.4\%, and 11.8\% MMLU, HumanEval, and GSM8K, respectively. This improvement arises because, in TokCom, the LLM does not need to generate a complex reasoning process from scratch; it only verifies and refines the simpler draft provided by the SLM. We also evaluate the communication overhead of TokCom, as shown in Fig. \ref{Experiment}(c). On HumanEval, the SLM transmits an average of 400 tokens to the LLM per correctly solved problem. These communication tokens incur no additional cost, as they correspond to the SLM's inherent reasoning output and would be produced regardless of whether they are transmitted. In resource-constrained scenarios, SLM-Only suffers from limited inference quality, whereas LLM-Only incurs a high computational cost. Our TokCom avoids both extremes and achieves a balance. It approaches the accuracy of LLM-Only with only a modest communication overhead, while significantly improving the overall computational efficiency of the system.

% Finally, we evaluate the communication overhead in TokCom. In HumanEval, the SLM transmits an average of 400 tokens to the LLM per correctly solved problem. Notably, these 400 communication tokens are not an independent cost, as they correspond to the SLM's inherent reasoning output and would be produced regardless of whether they are transmitted. From a system perspective, TokCom does not introduce any redundant reasoning generation but only transfers the draft for further verification. In resource-constrained scenarios, this design is particularly necessary because SLM-Only suffers from limited inference quality, while LLM-Only is too computationally demanding for edge devices. Therefore, TokCom achieves a balance. It approaches the accuracy of LLM-Only with only a modest communication overhead, while significantly improving the overall system computational efficiency.

\section{Conclusion and Future Directions}
In this section, we first present the key conclusions of this article and then elaborate on promising future research directions toward the practical implementation of TokCom in 6G networks.
\subsection{Conclusion}
TokCom represents a significant shift in the design and role of communication networks. By aligning network representations with the token-based abstractions used in modern AI systems, TokCom enables new forms of collaboration and efficiency that are difficult to achieve with traditional architectures. Applications range from generative recovery of missing or corrupted information to low-latency synchronization of distributed robotic systems, positioning TokCom as a candidate architectural blueprint for the emerging agentic computing era. Looking toward 6G and beyond, this perspective suggests a transition from networks that primarily deliver bits to networks that effectively deliver machine-interpretable intelligence. In this context, communication is no longer a separate utility but becomes an integral component of distributed computation. Despite its promise, implementing TokCom in practice requires addressing several technical and architectural challenges. The following subsections outline key directions for future research in this area.
\subsection{Open Research Challenges}
\subsubsection{Token Consensus in Dynamic Multi-Agent Networks}
The case study presented in Section III operates under a draft-and-verify paradigm between a specific SLM and an LLM. However, scaling TokCom to more dynamic, decentralized systems, such as autonomous-driving networks or multi-robot swarms, poses uncharted theoretical and architectural challenges. In distributed multi-agent systems, agents may have diverse functional specializations, varying computational capabilities, and different willingness to cooperate. Therefore, future research should move beyond simple pairs of models toward a generalized multi-agent token consensus framework that dynamically extracts the optimal functionalities of each agent. A primary challenge in this setup is identifying the reinforcing entity responsible for orchestrating token validation and mitigating semantic drift. For instance, in high-mobility C-V2X scenarios, relying solely on peer-to-peer voting introduces severe latency and risks token divergence. To address this, edge network infrastructure, such as roadside units and base stations, must serve as architectural anchors. Instead of executing full model reasoning, these edge nodes can act as semantic arbitrators, maintaining global context matrices, aggregating token drafts from surrounding vehicles, and coordinating lightweight voting consensus. Additionally, since real-world agents are unlikely to allocate computational resources for token verification without compensation, future extensions should incorporate token-based incentive schemes and game-theoretic mechanisms to motivate participation.
\subsubsection{Token-Native Protocols}  
Current networking protocols are largely agnostic to the semantic value of the data they transport and treat all bits equally important. In TokCom, however, different tokens can have vastly different impacts on downstream reasoning and control, for example, safety-critical decision tokens versus low-importance filler tokens. Therefore, a token-native protocol must enable the network to recognize, classify, and prioritize critical reasoning tokens with higher reliability and lower latency, while reducing the requirements for less important tokens. Achieving this requires rethinking the entire 6G protocol stack: the physical and MAC layers must support token-aware scheduling and protection; transport and network layers must provide token-level QoS differentiation; and the application layer must expose token importance and task semantics to the lower layers in a standardized manner.

\subsubsection{Security, Privacy, and Safety in TokCom}  
Because tokens are compact, information-dense representations, their interception or manipulation can expose sensitive information about a model’s internal state, user intent, or private context. TokCom therefore requires new mechanisms for secure and privacy-preserving token exchange. These include, for example, privacy-preserving token embeddings that obfuscate or disentangle sensitive attributes while preserving task utility, and federated token learning schemes that support collaborative model improvement without sharing raw data or full model parameters. 
In parallel, the use of generative repair mechanisms (e.g., token inpainting) introduces the risk of token hallucinations, in which plausible but incorrect or unsafe tokens are generated by the network or edge models. In safety-critical applications such as autonomous driving or industrial automation, such hallucinations can lead to catastrophic consequences. Thereby, robust detection, calibration, and verification of generatively reconstructed tokens are essential components of any practical TokCom architecture.

\subsubsection{Standardization and Semantic KPIs}
Realizing TokCom  requires a concerted standardization effort within organizations such as 3GPP, IEEE, and ITU. To ensure interoperability across devices, vendors, and operators, standardized token encoders and decoders, as well as interfaces, must be defined, specifying how tokens are represented, compressed, protected, and exchanged consistently. Conventional communication metrics such as bit error rate and packet delivery ratio must be refined to capture AI-centric performance, including knowledge accuracy (e.g., how faithfully the transmitted tokens preserve intended semantics), task success rate (e.g., end-to-end effectiveness on target AI tasks), and reasoning latency (e.g., the delay from token exchange to actionable inference). As the industry progresses toward AI-native radio access networks, the integration of TokCom into these standards, together with agreed reference architectures and evaluation methodologies, will be a defining characteristic of next-generation connectivity.

\normalem
\bibliographystyle{IEEEtran}  % 设置IEEE样式
\bibliography{Ref}  % 加载参考文献数据库

\end{document}